\documentclass[10pt,a4paper]{article}
\usepackage[latin1]{inputenc}
\usepackage{amsmath}
\usepackage{amsfonts}
\usepackage{amssymb}
\usepackage{epsfig}
\usepackage[lmargin=2cm,rmargin=2cm,tmargin=2cm,bmargin=2cm]{geometry}

\begin{document}

\title{Generalized Uncertainty Principle, Modified Dispersion Relation\\ and Barrier penetration by a Dirac particle}
\author{{\bf Sumit Ghosh} \thanks{E-mail: sumit.ghosh@bose.res.in}
 \\ S. N. Bose National Centre for Basic Sciences \\ Block JD, Sector III, Salt Lake, Kolkata 700098}

\maketitle

\begin{abstract}
We have studied the energy band structure of a Dirac particle in presence of a generalised uncertainty principle (GUP). We start from defining a modified momentum operator and derive corresponding modified dispersion relation (MDR) and GUP. Apart from the forbidden band within the range $\pm m$, $m$ being the mass of the particle, we find the existence of additional forbidden bands at the both ends of the spectrum. Such band structure forbids a Dirac particle to penetrate a potential step of sufficient height ($\sim E_P$, $E_P$ being Planck energy). This is also true for massless particle. Unlike the relativistic case, a massless particle also can reflect from a barrier of sufficient height. Finally we discuss about the Klein's paradox in presence of the GUP. 

\end{abstract}

\section{Introduction}
Introduction of a minimum length scale \cite{aml}-\cite{panes} has revealed several salient features of gravity and quantum field theory. The essence of a minimum length can be captured by various means - by a generalized uncertainty principle (GUP), by a modified dispersion relation (MDR), by  a deformed (doubly) special relativity (DSR) etc. One can initially start from a modified momentum operator and successively obtain a GUP or an MDR \cite{hoss3}. The manifestation of the presence of a minimal length in different fields, such as black hole thermodynamics \cite{aml2} -\cite{me}, relativistic and non-relativistic quantum mechanics \cite{hoss4}-\cite{lew}, quantum space time phenomenology \cite{camlr} etc, has stimulated a large number of extensive studies.

In this paper we study the energy band structure of a Dirac particle in presence of a GUP. A GUP is associated with a minimum measurable length (and a maximum energy in some cases). We will see how their existence is manifested when a relativistic particle tries to penetrate a potential wall. We thoroughly analysed the MDR which we derive from a modified momentum operator consistent with the GUP. We find that the allowed energy levels are not extended to $\pm \infty$, rather confined within a finite region. Such band structure gives rise to unique features to the barrier penetration. In standard relativistic theory we know that for barrier height $V>2m$, the incoming particle may propagate through the potential region. This is the well known {\it Klein's paradox} \cite{scbl}. The phenomenon is now well understood in terms of many particle \cite{rav,hol,dom} as well as one particle theory \cite{alh}. In presence of a GUP we find that a particle can't penetrate a potential region if its height exceeds certain upper limit which is of the order of Plack energy ($E_P$). We thoroughly study the conditions of tunnelling for both massive and massless particle and also discuss about the Klein's paradox.   

The organisation of this paper is as follows. In section 2, We introduce a modified momentum operator and obtain the corresponding Dirac equation with a discussion on the structure of the  modified commutation relation and GUP. In section 3, we find the solution for the modified Dirac equation and derive the MDR which is used to find the band structure. This band structure is used to study the nature of barrier penetration for both massive and massless particles in section 4. In section 5 we discuss about the Klein's paradox. The conclusion on the whole work is kept in section 6.


\section{GUP and the Dirac equation}
In this section we start from a modified momentum operator and derive corresponding GUP and the modified Dirac equation.

A GUP corresponds a modified commutation relation between position and momentum. One way to achieve this is to define a modified momentum operator whose structure is determined by the GUP. Here we use a GUP which contains both odd and even powered terms in momentum. This kind of GUP is recently proposed in \cite{sdas}. It not only accommodates a minimum length, but also a maximum energy. We show a systematic derivation of the GUP starting from a modified momentum operator and consecutively obtain the modified Dirac equation.

We start by defining a new momentum operator($\vec{P}$), keeping first two correction terms
\begin{eqnarray}
P_i = p_i \left( 1-a_0 \frac{p}{E_P/c} + b_0 \frac{p^2}{(E_P/c)^2} \right)
\label{P0}
\end{eqnarray}
where the suffix $i$ ($i=1,2,3$) corresponds the $i-th$ component. $E_P$ is the Planck energy. $a_0$ and $b_0$ are real positive numbers. $p$ is the magnitude of the operator $\Sigma p_i^2$. $p_i$ obeys the standard commutation relation
\begin{eqnarray}
[x_j,p_i]=i\hbar \delta_{ji}
\label{xp}
\end{eqnarray}
$x_j$ is the $j-th$ component of position operator.

It would be convenient for future calculations to introduce the scaled coefficients
\begin{eqnarray}
a=\frac{a_0}{E_P/c} \hspace{1cm} b=\frac{b_0}{(E_P/c)^2}
\label{ab}
\end{eqnarray}
and express (\ref{P0}) in terms of them. The modified momentum operator can be recast as 
\begin{eqnarray}
P_i = p_i(1-a p + b p^2)
\label{p3}
\end{eqnarray}
We also define the new position operator as
\begin{eqnarray}
\vec{X} = \vec{x}
\label{x}
\end{eqnarray}

These modified momentum ($P$) and position operator ($X$) satisfy the commutation relation 
\begin{eqnarray}
[X_i,P_j] = i\hbar \left( \delta_{ij} - a \left( \delta_{ij} p + \frac{p_i p_j}{p}\right) + b \left( \delta_{ij} p^2 + 2p_ip_j \right)\right)
\label{com} 
\end{eqnarray}

We shall now obtain the modified Dirac equation. The Dirac equation in presence of the modified momentum operator (\ref{p3}) is given by \cite{sdas2,ped}
\begin{eqnarray}
H \psi = \left[ c(\vec{\alpha} \cdot \vec{p})(1-a (\vec{\alpha} \cdot \vec{p}) + b (\vec{\alpha} \cdot \vec{p})(\vec{\alpha} \cdot \vec{p})) + \beta m c^2 \right]\psi
\label{md}
\end{eqnarray}
where we have used Dirac's prescription for the operator $p$ (i.e $p\rightarrow \vec{\alpha} \cdot \vec{p}$) \cite{sdas2,ped}. $\alpha$ and $\beta$ are the Dirac matrices defined as
\begin{eqnarray}
\alpha_i = \begin{pmatrix}0&\sigma_i \\ \sigma_i & 0 \end{pmatrix} , \hspace{2cm} \beta = \begin{pmatrix} I & 0 \\ 0 & -I \end{pmatrix}
\end{eqnarray}
where $\sigma_i$ is the $ith$ Pauli matrix ($i$ = 1, 2, 3).

For simplicity we will consider one dimensional motion (say along z axis). In one dimension the modified momentum operator (\ref{p3}) and the commutation relation (\ref{com}) becomes
\begin{eqnarray}
P &=& p(1+ap+bp^2) \label{p} \\
{[X,P]} &=& i\hbar \left( 1 - 2aP + (3b-2a^2)P^2 \right) \
\label{com2} 
\end{eqnarray}

The GUP in one dimension in terms of the new operators $P$ and $X$ is
\begin{eqnarray}
\Delta X \Delta P &\geq & \vert \frac{1}{2} \left\langle [X,P] \right\rangle \vert \nonumber \\
&\geq & \frac{\hbar}{2} (1-2 a \left< P \right> + (3 b-2a^2) \left< P^2 \right>)
\label{gup0}
\end{eqnarray}

Note that by putting $b = 2a^2$ we can obtain the GUP proposed in \cite{sdas}.

For minimum position uncertainty, we can replace $\left<P\right>$ and $\left<P^2\right>$ by $\Delta P$ and  $(\Delta P)^2$ \cite{me} and hence the GUP becomes
\begin{eqnarray}
\Delta X \Delta P \geq \frac{\hbar}{2} (1-2 a\Delta P + (3 b-2a^2) (\Delta P)^2)
\label{gup}
\end{eqnarray}

The existence of minimum position uncertainty (i.e. $\Delta X$ to have a minima) requires the coefficient of $(\Delta P)^2$ to be positive. i.e. ($3b-2a^2 > 0$), and for the minimum length to be positive we need
\begin{eqnarray}
b > a^2
\label{con1}
\end{eqnarray}
In \cite{sdas}, they have used $b = 2a^2$ which is also consistent with the above condition (\ref{con1}).

For a particle moving along $z$ direction (\ref{md}) simplifies as 
\begin{eqnarray}
\left[c \alpha_z p - ca p^2 + c b \alpha_z p^3 + \beta m c^2 \right] \psi = E_G \psi 
\label{mDe1}
\end{eqnarray}
where $E_G$ is the GUP corrected energy of the system. For further calculations we will replace the momentum operator $p$ by its differential form $-i\hbar \frac{d} {d z}$ and recast the modified Dirac equation as
\begin{eqnarray}
\left[-i \hbar c \alpha_z \frac{d} {d z} +\hbar^2 ca \frac{d^2} {d z^2} -i \hbar c b \alpha_z \frac{d^3} {d z^3} + \beta m c^2 \right] \psi = E_G \psi 
\label{mDe}
\end{eqnarray}

Note that the modified Dirac equation (\ref{mDe}) contains third order spatial derivative and an exact solution requires the continuity of higher derivatives also. Since the GUP correction is very small one can also take a perturbative approach \cite{vag}-\cite{ped} as well. 


\section{Solution of modified Dirac equation for a free particle and its band structure}

The standard one dimensional Dirac equation for a free particle is given by
\begin{equation}
\left(-i\hbar c \alpha_z \frac{d}{dz} + \beta m c^2 \right) \psi(z) = E \psi(z)
\end{equation}
Note that one can also obtain this equation by putting $a=b=0$ in (\ref{mDe}). Its positive energy (+E) solution is given by
\begin{equation}
\psi(z) = N e^{ikz} \begin{pmatrix} \chi \\ \mathcal{U} \sigma_z \chi \end{pmatrix}
\label{psi}
\end{equation}
Here $\chi$ is a $(2\times 1)$ column matrix which satisfies orthonormality condition $\chi^\dagger \chi =1$, $N$ is the normalization factor and $\mathcal{U}$ is a dimensionless quantity defined as
\begin{eqnarray}
\mathcal{U} = \frac{c \hbar k}{E + mc^2}
\label{u}
\end{eqnarray}
The wave number ($k$) obeys the dispersion relation
\begin{eqnarray}
E^2 = (c\hbar k)^2 + m^2 c^4
\label{dr}
\end{eqnarray}   

For the modified Dirac equation (\ref{mDe}), the solution would not be so simple. Whatever the solution be, it must converge to (\ref{psi}) in absence of GUP. Let us consider the following ansatz 
\begin{eqnarray}
\psi_G(z) = N e^{ik_Gz} \begin{pmatrix} \chi \\ \mathcal{U}_G \sigma_z \chi \end{pmatrix}
\label{psig}
\end{eqnarray} 
to be a solution of (\ref{mDe}).

The assumption
\begin{eqnarray}
\lim_{(a,b) \rightarrow 0} \psi_G \rightarrow \psi
\end{eqnarray} 
where $\psi$ is given by (\ref{psi}), implies
\begin{eqnarray}
&& \lim_{(a,b) \rightarrow 0} \mathcal{U}_G \rightarrow \mathcal{U}
\label{anzU}\\
&& \lim_{(a,b) \rightarrow 0} E_G^2 \rightarrow (c\hbar k_G)^2 + m^2 c^4
\label{anzE}
\end{eqnarray}
where $\mathcal{U}$ is given by (\ref{u}).
We now substitute ansatz (\ref{psig}) in (\ref{mDe}) and obtain the GUP corrected expression for energy ($E_G$) or the MDR and $\mathcal{U}_G$ as
\begin{eqnarray}
E_G &=& \pm \sqrt{c^2\hbar^2 k_G^2 (1+b\hbar^2 k_G^2)^2 + m^2 c^4} - ca\hbar^2 k_G^2 
\label{mdr}\\
\mathcal{U}_G &=& \frac{c \hbar k_G (1+b\hbar k_G^2)}{ca\hbar^2 k_G^2 + mc^2 + E_G}
\label{ug}
\end{eqnarray}
One can readily see that both $E_G$ and $\mathcal{U}_G$ satisfy (\ref{anzE}) and (\ref{anzU}). 
 
The condition for the existence of a positive energy solution is
\begin{eqnarray}
\sqrt{c^2\hbar^2 k_G^2 (1+b\hbar^2 k_G^2)^2 + m^2 c^4} > ca\hbar^2 k_G^2
\label{pcon}
\end{eqnarray}
If we consider particles for which $\hbar k>>m$ and $0<b(\hbar k_G)^2<<1$ then by using (\ref{con1}) the condition (\ref{pcon}) can simplified to
\begin{eqnarray}
(a^2-2b)< 0
\label{con2}
\end{eqnarray}
Notice that in \cite{sdas}, $b=2a^2$ which also satisfies this condition.

By keeping our interest confined within terms $\mathcal{O}(a^2,b)$, the expression for the wave vector $k_G$ can be obtained from (\ref{mdr}) as
\begin{eqnarray}
\hbar^2 k_G^2 = \frac{E_G^2-m^2c^4}{c^2 - 2aE_Gc} \left[1 + \frac{(a^2-2b)(E_G^2-m^2c^4)}{(c-2aE_G)^2} \right]
\label{k}
\end{eqnarray}
which is also consistent with (\ref{anzE}). 

Let us discuss some properties of (\ref{k}) and let see what they say about the GUP (\ref{gup}). Let us first consider the denominator $(c-2aE_G)$. Using the basic definition (\ref{ab}) we can rewrite this as $(1-2a_0 \frac{E_G}{E_P})$. It clearly shows that as the particle energy $E_G$ approaches $E_P$ the wave vector ($k_G$) tends to a divergence and for $E_G>E_P/2a$, $k_G \rightarrow \pm i\infty$. Hence it predicts a upper limit for energy which can be achieved by a particle and is given by
\begin{eqnarray}
E_G\vert_{max} = \frac{E_P}{2a_0}
\label{emax}
\end{eqnarray} 
Note that the order of maximum energy matches well with that predicted in \cite{sdas}.

\begin{figure}[h]
\centering
\epsfig{file=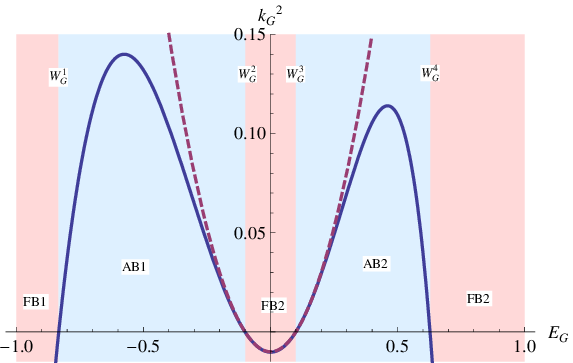,scale=.8}
\caption{$k_G^2$ as a function of energy $E_G$ for ($a=0.1$, $b=1$, $m=0.1$)  with $c = \hbar = E_P =L_P = 1$. The dashed line shows non GUP dispersion relation $k^2 = E^2 - m^2$. The light-red regions (FB1, FB2, FB3) are forbidden bands and the light-blue regions (AB1, AB2) are the allowed bands.}
\label{k2e}
\end{figure}

We will now study the band structure. The simplest way to know the occurrence of a propagative or non-propagative mode is to find the nature of $k_G^2$ with respect to energy ($E_G$). A negative value of $k_G^2$ suggests a imaginary  $k_G$ which is a damped mode. Corresponding energy values form a forbidden band. In $fig.$\ref{k2e} we plot $k_G^2$ as a function of $E_G$ as given by equation (\ref{k}), which shows the band structure for the modified Dirac equation. The forbidden band within the range $\pm m$ is well known in Dirac theory without GUP. The interesting feature is the presence of the other forbidden bands which are the sole effect of the GUP. The edges of the allowed and forbidden bands, which are actually zeros of (\ref{k}), are given by
\begin{eqnarray}
W_G^1 &=& - \frac{\sqrt{c^2 \left(2b-a^2\right) \left(1+ \left(2b-5 a^2\right)m^2 c^2\right)}+2ac}{2 b-5a^2}
\label{eg1}\\
W_G^2 &=& -m
\label{eg2}\\
W_G^3 &=& m
\label{eg3}\\
W_G^4 &=& \frac{\sqrt{c^2 \left(2b-a^2\right) \left(1+ \left(2b-5 a^2\right)m^2 c^2\right)}-2ac}{2 b-5a^2}
\label{eg4}
\end{eqnarray}
One can see that for $2b\rightarrow5a^2$, $W_G^1$ diverges whereas $W_G^4$ still remain finite. For $2b<5a^2$ one can check that $W_G^1>E_G\vert_{max}$ whereas for $2b>5a^2$, both $W_G^1$ and $W_G^4$ are finite. The edge $W_G^1$ is the bottom of the Dirac sea and $W_G^4$ gives the maximum positive energy for a particle. In the following sections we will see how this boundedness of the negative energy gives rise to unique features of barrier penetration. In \cite{sdas}, $b=2a^2$ and that is why the boundedness of the negative energy is not there. From now on we will work within the region $2b>5a^2$. Based on the existence of these four edges, we can divide the entire energy spectrum in five different regimes which are described in $Table$\ref{band}.

\begin{table}[h]
\begin{center}
\begin{tabular}{|c|c|c|}
\hline
FB1 & 1st forbidden band & $ E_G < W_G^1 $\\
\hline
AB1 & 1st allowed band & $ W_G^2 < E_G < W_G^1 $\\
\hline
FB2 & 2nd forbidden band & $ W_G^3 < E_G < W_G^2$\\
\hline
AB2 & 2nd Allowed band & $ W_G^4 < E_G < W_G^3 $\\
\hline
AB3 & 3rd forbidden band & $ W_G^4 < E_G  $\\
\hline
\end{tabular}
\end{center}
\caption{Nomenclature and positions of different bands}
\label{band}
\end{table}

\begin{figure}[h]
\centering
\epsfig{file=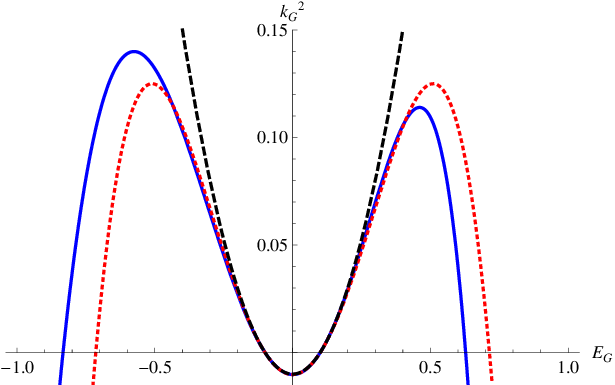,scale=.8}
\caption{Variation of $k_G^2$ with $E_G$ for [$a=0.1,b=1.0$] (blue line), [$a=0,b=1.0$] (red dotted line) and [$a=0,b=0$] for $m=0.1$ ($c=\hbar=1$). The blue line corresponds the GUP (\ref{gup}) we used, red dotted line corresponds the GUP with even powers of momentum \cite{scar,me} and the black dashed line shows the standard relativistic dispersion relation.}
\label{ek0}
\end{figure}

Let us now focus on the dispersion curve ($fig.$\ref{k2e}). The dashed line in $fig.$\ref{k2e} corresponds the standard relativistic dispersion relation and the solid line gives the GUP corrected MDR. We see that for low energy the solid line matches well with the relativistic dispersion relation. As $E_G$ is increased, $k_G^2$ gradually attains a maximum value and then start decreasing. The asymmetry of the curve is due to term $a$. If we put $a=0$ then the curve is symmetric\footnote{In that case the GUP (\ref{gup}) is $\Delta P \Delta X \geq \frac{\hbar}{2} (1+3b(\Delta P)^2)$, which is similar to that predicted in string theory \cite{scar}. It supports the existence of a minimum length scale but doesn't suggest a maximum energy. It agrees with the fact that for $a=0$, $E_G|_{max}=\infty $ (\ref{emax}). The energy levels are bounded in both ends as it also satisfies $2b>5a^2$.}
about the $k_G^2$ axis ($fig.$\ref{ek0}). The upper bound of the $k_G^2$ suggests the fact no matter how energetic particle we use we can't reduce its wave length below a certain limit and hence we can't measure below a certain length. 

With this let us now proceed to the next section where we will discuss about barrier penetration in this modified picture.

\section{Effect of GUP and MDR on barrier penetration}

Let consider a particle incident on a potential step of height $V_0$. According to nonrelativistic quantum mechanics the particle can't penetrate inside the step if $V_0$ is more than the particle energy ($E$). There will be no propagating mode inside the potential region and the motion will be exponentially damped. In relativistic approach, if $V_0>2m$, then a particle with energy $m<E<V_0-m$ can propagate inside the potential region. This is known as $Klein's~paradox$ \cite{scbl} which will be discussed in next section. In this section we will see what the situation will be when a particle encounters a potential step in presence of a GUP.

\begin{figure}[h]
\centering
\epsfig{file=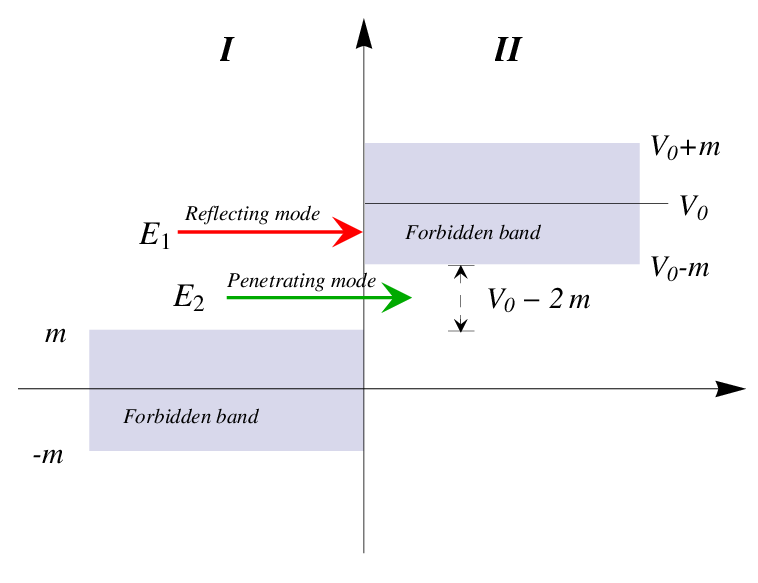,scale=0.75}
\caption{Two free particles with energy $E_1$ and $E_2$ facing a potential step of height $V_0$ ($V_0>E_{1,2}$). The shaded regions are forbidden bands. A particle can not penetrate region (II) if it encounter a forbidden band there. Hence the particle with energy $E_1$ will bounce back while that with energy $E_2$ will propagate through the potential region. From the figure it is clear that the possibility of a propagating mode arises only for $V_0>2m$.}
\label{step}
\end{figure}

To determine the existence of a propagating mode we will use the same method as the previous section, i.e. to check the signature of the square of the wave vector. The expression can easily be found by replacing $E_G$ by $E_G-V_0$ in (\ref{mdr})and is given by     



\begin{eqnarray}
\hbar^2 q_G^2 &=& \frac{(E_G-V_0)^2-m^2c^4}{c^2 - 2a(E_G-V_0)c} \left[1 + \frac{(a^2-2b)((E_G-V_0)^2-m^2c^4)}{4(c-2a(E_G-V_0))^2} \right] 
\label{q}
\end{eqnarray}

In relativistic quantum mechanics so long $V_0>2m$ there will always be a propagating mode. The theory predicts a finite transmission even for $V_0 \rightarrow \infty$, which is also a manifestation of Klein's paradox. It happens because the allowed energy bands are unbounded and for $V_0>2m$ the potential can elevate an allowed negative energy state to the level of a positive energy state. But in presence of GUP the allowed bands are localised within certain region. This in turns restricts the transmission process.

\begin{figure}[h]
\centering
\epsfig{file=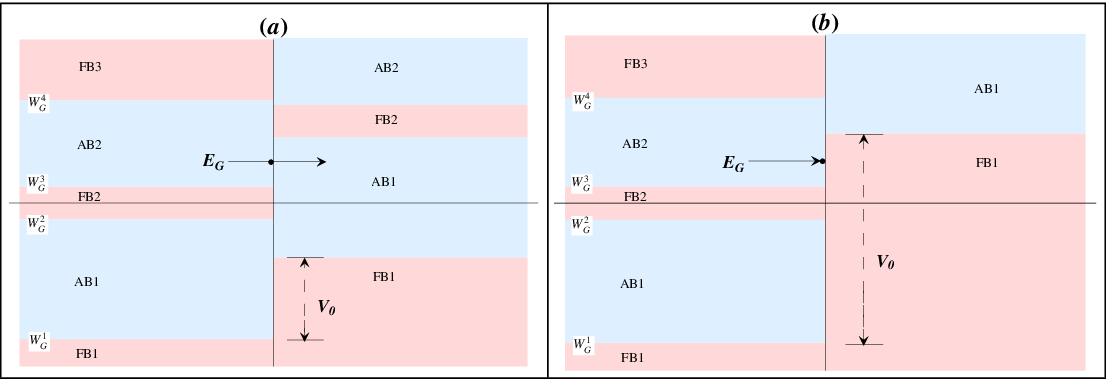,scale=.75}
\caption{Energy band structure inside and outside the potential region with a=0.05, b=1 and m=.25. We have shown the situation for two different values of applied potential ($a$) $V_0=1.3$ and ($b$) $V_0=2.5$. We consider a particle with $E_G=0.5$. Although in both cases $V_0>2m$, we see that only in ($a$) the particle can penetrate in the potential region.}
\label{barrier}
\end{figure}

We see from $fig.$\ref{k2e} there are two allowed bands ($AB1$ and $AB2$) which correspond  propagating modes. The only chance of penetration for a particle from $AB2$ with energy $E_G<V_0$ appears if $V_0$ can raise an energy level from $AB1$ to the level of $E_G$ in the potential region. From $fig.$\ref{step} we see that it is possible only when $V_0>2m$. All the particles with energy $m<E<V_0-m$ can propagate inside the potential ($fig.$\ref{barrier}$a$). But if we keep increasing $V_0$, gradually it will bring $FB1$ band to this level ($fig.$\ref{barrier}$b$) and the particle will again be prohibited to enter. From $fig.$\ref{barrier}($a,b$) it is clear that if $V_0>(W^G_4+|W^G_1|)$, where $W^G_1$ and $W^G_4$ are the extreme edges of $AB1$ and $AB2$ (\ref{eg1},\ref{eg4}), The there will be no possibility of transmission. This upper limit of the potential is given by
\begin{eqnarray}
V_{max} = W_G^4+|W_G^1| = \frac{2 c \sqrt{\left(2 b-a^2\right) \left(c^2 m^2 \left(2 b-5 a^2\right)+1\right)}}{2 b-5 a^2}
\label{vmx}
\end{eqnarray}   
Thus we see that for a massive particle with $E_G<V$ tunnelling is possible only for a finite window ($m<V<V_{max}$). In the next subsection we will see the scenario for a massless particle.

\subsection{Barrier penetration by a massless particle}
\begin{figure}[h]
\centering
\epsfig{file=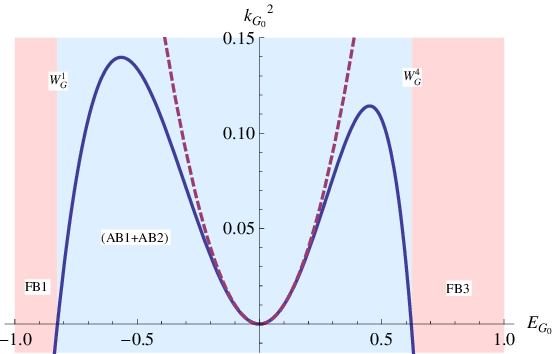, scale=0.8}
\caption{$k_{G_0}^2$ as a function of energy for a massless Dirac particle for $a=0.1$ and $b=1$. There is no forbidden energy band around the zero energy, but the forbidden energy bands on the both ends still exist (FB1 and FB3). The dashed line shows the relativistic dispersion relation in absence of GUP.}
\label{vq0}
\end{figure}

According to standard relativistic theory a massless particle can penetrate a potential barrier of any height. The tunnelling starts from $V=0$. This will be more clear if we look at the expression of its wave vector. For a massless particle the wave vector can easily be obtained by putting $m=0$ in (\ref{k}) and is given by
\begin{eqnarray}
\hbar^2 k_{G_0}^2 = \frac{E_{G_0}^2}{c^2 - 2aE_{G_0}c} \left[1 + \frac{(a^2-2b)E_{G_0}^2}{(c-2aE_{G_0})^2} \right]
\label{k0}
\end{eqnarray} 
One can readily see from (\ref{k0}) that $k_{G_0}$ has only two non-zero roots and rest of the two roots are at $E_G=0$. This suggests that there will be only three bands. The positions of the bands can be better understood if we look at the dispersion curve for massless particle ($fig.$\ref{vq0}). 

From $fig.$\ref{vq0} we see that the forbidden band $FB2$ is absent but the other two forbidden bands $FB1$ and $FB3$ are still present.The location of band edges also can be found from (\ref{eg1}-\ref{eg4}) by putting $m=0$. The maximum value of potential which allows a penetrating mode for a massless particle is thus given by
\begin{eqnarray}
V_{max_0} =(W_G^4+|W_G^1|)_{m=0} = \frac{2 c \sqrt{\left(2 b-a^2\right)}}{2 b-5 a^2}
\label{vmx0}
\end{eqnarray}

The existence of the forbidden bands at two extremes ($FB1,FB3$) is solely due to the GUP correction. According to standard relativistic theory, a massless particle can not be stopped by any potential step, as there is no forbidden band. But with GUP correction we see that a massless particle can not pass through a potential barrier with hight $V_0 > V_{max_0}$. Thus reflection of a massless particle from a potential step will be the evidence for the existence of a minimum length manifested here through a modified momentum operator (\ref{p}).  
 
\section{GUP, MDR and Klein's paradox}
Klein's paradox is a well known phenomenon in relativistic quantum theory. When a relativistic particle with energy $m<E<V_0-m$ incident on a potential step with $V_0>2m$, the reflectivity ($R$) becomes greater than 1 and consequently the transmissivity ($T$) is negative. The paradox is resolved and is well understood in terms of pair production by the barrier \cite{rav,hol,dom}. We will not go through the details of the paradox and its resolution. Instead we will keep our discussion confined within the occurrence and nature of the paradox. 

in standard relativistic theory the solution of Dirac equation for a free particle incident on a potential step $V_0 \Theta(0)$ ($\Theta$ here is the Heaviside step function) can be written as
\begin{eqnarray}
\psi(z) &=& A \begin{pmatrix}\chi \\ \mathcal{U} \sigma_z \chi \end{pmatrix}e^{ikz} + B \begin{pmatrix}\chi \\ -\mathcal{U} \sigma_z \chi \end{pmatrix}e^{-ikz} \hspace{1cm}z<0 \nonumber \\
&=& C \begin{pmatrix}\chi \\ \mathcal{U'} \sigma_z \chi \end{pmatrix}e^{iqz} \hspace{4.1cm}z>0
\end{eqnarray}  
where $A$,$B$ and $C$ are undetermined constants. $k=\sqrt{E^2-m^2c^4}/\hbar c$ and $q=\sqrt{(E-V_0)^2-m^2c^4}/\hbar c$. $\mathcal{U}$ is given by (\ref{u}) and $\mathcal{U}'$ can be obtained from (\ref{u}) by replacing $E$ with $E-V_0$. The reflectivity ($R$) and transmissivity ($T$) is given by \cite{scbl,rav,hol,dom}
\begin{eqnarray}
R = \left( \frac{1-\kappa}{1+\kappa} \right)^2 \hspace{2cm} T = \frac{4\kappa}{(1+\kappa)^2}
\end{eqnarray} 
where $\kappa$ is given by
\begin{eqnarray}
\kappa = \frac{\mathcal{U'}}{\mathcal{U}} = \frac{q}{k} \frac{E+m}{E-V_0+m} 
\label{kappa}
\end{eqnarray}
One can easily see that for $E<V_0-m$, $\kappa$ is negative which makes $R>1$ and $T<0$. This is Klein's paradox. The paradox is resolved by taking pair production by the barrier into account \cite{rav,hol,dom}. The negative $T$ is due to the presence of antiparticles. The probability of pair production is given by 
\begin{eqnarray}
\mathcal{P} = -\frac{4\kappa}{(1-\kappa)^2}
\label{pair}
\end{eqnarray} 

Now we will study the situation in presence of a GUP. We can start from ansatz (\ref{psig}) which will give the following expressions for reflectivity ($R_G$) and transmissivity ($T_G$).
\begin{eqnarray}
R_G = \left( \frac{1-\kappa_G}{1+\kappa_G} \right)^2 \hspace{2cm} T_G = \frac{4\kappa_G}{(1+\kappa_G)^2}
\label{rg}
\end{eqnarray}
where $\kappa_G$ is given by
\begin{eqnarray}
\kappa_G = \frac{\mathcal{U'}_G}{\mathcal{U}_G} = \frac{q_G (1+b\hbar q_G^2)}{k_G (1+b\hbar k_G^2)} \frac{ca\hbar^2 k_G^2 + mc^2 + E_G}{ca\hbar^2 q_G^2 + mc^2 + (E_G-V_0)}
\label{kappaG}
\end{eqnarray}
where $k_G$ and $q_G$ are given by (\ref{k}) and (\ref{q}). Similarly the pair production probability is given by
\begin{eqnarray}
\mathcal{P}_G = -\frac{4\kappa_G}{(1-\kappa_G)^2}
\label{pairG}
\end{eqnarray} 

\begin{figure}[h]
\centering
\epsfig{file=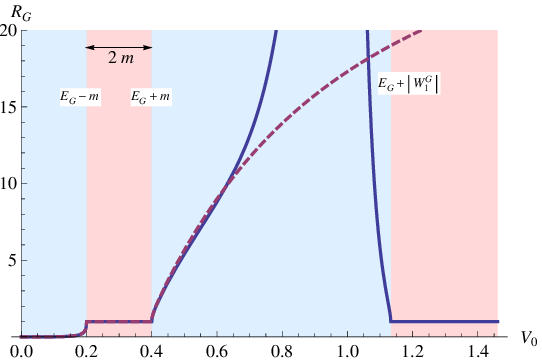,scale=.8}
\caption{Reflectivity ($R_G$) as a function of the potential height ($V_0$) for a particle with $m=0.1,~E_G=0.3$ and $a=0.1,~b=1$ ($c=\hbar=1$). The solid line shows the GUP corrected value and the dashed line shows the standard relativistic result. $R_G=1$ for $E_G-m<V_0<E_G+m$ (particle facing $FB2$) and for $V_0>E_G+W_1^G$ (particle facing $FB1$) whereas without GUP correction reflectivity keeps increasing as $V_0>E_G+m$.}
\label{refl}
\end{figure}

Klein paradox is manifested by a reflectivity higher than 1 and its occurrence can be detected by observing the nature of $R_G$ which is shown in $fig.$\ref{refl}.

\begin{figure}[h]
\centering
\epsfig{file=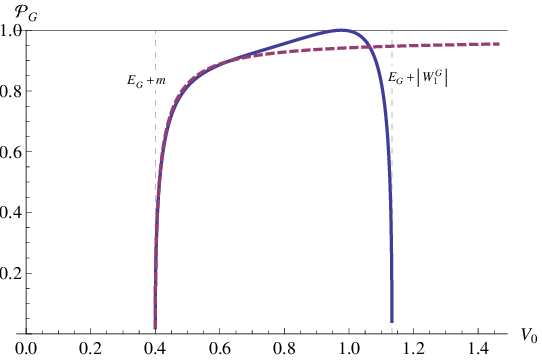,scale=1}
\caption{Variation of $\mathcal{P}_G$ with $V_0$ for a particle with $m=0.1,~E_G=0.3$ with $a=0.1,~b=1$. Solid line shows the GUP corrected value and the dashed line is the nonGUP result. According to non GUP prediction pair production rate will keep increasing with energy and finally reach unity as $E \rightarrow \infty$. In presence of GUP $\mathcal{P}_G$ initially increases a similar way and then becomes unity and finally diminish to zero as $V_0$ exceeds certain upper limit.}
\label{pp}
\end{figure}

In $fig.$\ref{refl} we see that $R_G$ starts from almost zero value and becomes unity at $V_0=E_G-m$. This is when the particle first faces $FB2$ in the potential region. $R_G$ remain unity till $V_0=E_G-m$. The particle then faces $AB1$ in the potential region. But instead of decreasing $R_G$ starts increasing in this region. This is because the barrier penetration is associated with pair production in this region. We see that so far the result matches well with the standard relativistic result. But as we make $V_0>E_G+|W_1^G|$ we see that $R_G$ again reduces to unity. This is when the particle faces $FB1$ and can't penetrate any more. The situation will be more clear if we look at the change of the pair production probability ($\mathcal{P}_G$) with $V_0$ ($fig.$\ref{pp}).

From $fig.$\ref{pp} we see that  $\mathcal{P}_G$ is zero for both $V_0<E_G+m$ and $V_0>E_G+|W_1^G|$ and is finite in between, i.e. when the particle faces $AB1$. For $V_0<E_G+m$ the potential is not strong enough to produce a pair whereas for $V_0>E_G+|W_1^G|$ the potential corresponds an energy state from $FB1$ which prevents the pair production. $\mathcal{P}_G$ becomes unity at some point before its right end. The corresponding $V_0$ can be evaluated by putting $\kappa_G=-1$ (\ref{pairG})\footnote{One can see in $fig.$\ref{refl} that at that point $R_G$ diverges which also can be verified by putting $\kappa_G=-1$ in (\ref{rg}). For a non-GUP case this happens for $V_0\rightarrow\infty$.}. Thus we see that pair production takes place only for a finite range of $V_0$ and that is why barrier penetration also stopped as $V_0$ exceeds certain upper limit. Since the initial particle originated from $AB2$, its energy has an upper limit $W_4^G$. Hence the maximum value of $V_0$ for which pair production is possible is given by $W_4^G+|W_1^G|$, i.e $V_{max}$. Thus we see that in presence of a GUP the occurrence  of Klein's paradox is also limited within a finite window of the applied potential.


\section{Conclusion}

In this paper we study the MDR (\ref{mdr}) of a Dirac particle in presence of GUP (\ref{gup}) which accommodates a minimum length as well as a maximum energy. We see that within a certain range of the GUP parameters ($a,b$) the allowed energy states are bounded for both massive ($fig.$\ref{k2e}) and massless ($fig.$\ref{vq0}) particles. Due to such band structure  even a massless particle can fail to penetrate a potential barrier if its height ($V_0$) exceeds certain upper limit. The situation is further clarified by analysing the reflectivity ($R_G$) and pair production probability ($\mathcal{P}_G$). For small potential ($V_0<<E_G|_{max}$) $R_G$ ($fig.$\ref{refl}) and $\mathcal{P}_G$ ($fig.$\ref{pp}) match well with the standard relativistic predictions, but for higher energy they show completely different character. We show that above certain potential there are again complete reflection and zero particle production which suggests zero barrier penetration.

Such corrections however is far from being observed in an experiment. In a collision with centre of mass energy 1 TeV (LHC energy scale) the contribution of present GUP theories is 1 part in $10^{32}$. Besides, considering the GUP parameters to be of the order of unity one can see that the minimum potential required to stop a massless particle (roughly half of $V_{max_0}$(\ref{vmx0})) is of the order of Planck energy which is far beyond the reach of present technology. One possible ground where our findings can play an important role is in the vicinity of black holes. Their massiveness makes them suitable for using our findings to test the quantum gravity effects on the tunnelling mechanism \cite{mich} as well as on the  {\it Klein's paradox} like phenomena around them \cite{full}.

\section{Acknowledgement}
The author likes to thank Council of Scientific and Industrial Research (CSIR), India for financial support.


\end{document}